\documentclass{article}

\PassOptionsToPackage{numbers}{natbib}
\usepackage[final]{neurips_2024}
\usepackage[utf8]{inputenc} 
\usepackage[T1]{fontenc}    
\usepackage{hyperref}       
\usepackage{url}            
\usepackage{booktabs}       
\usepackage{amsfonts}       
\usepackage{nicefrac}       
\usepackage{microtype}      
\usepackage{xcolor}         
\usepackage{graphicx}
\usepackage{amsmath}
\usepackage{newtxtext,newtxmath}
\usepackage{float}
\usepackage{svg}

\newcommand{\llaves}[1]{\lbrace #1\rbrace}

\newcommand{\ket}[1]{| #1\rangle}

\title{Hybrid classical-quantum architecture for vectorised image classification of hand-written sketches}

 \author{
  Y. Cordero, \quad  S. Biswas, \quad F. Vilariño, \quad M. Bilkis \\
  Computer Vision Center (CVC), Barcelona, SPAIN. \\
  Universitat Autònoma de Barcelona (UAB), Barcelona, SPAIN.
 }

\begin{document}

\maketitle

\begin{abstract}
Quantum machine learning (QML) investigates how quantum phenomena can be exploited in order to learn data in an alternative way, \textit{e.g.} by means of a quantum computer. While recent results evidence that QML models can potentially surpass their classical counterparts' performance in specific tasks, quantum technology hardware is still unready to reach quantum advantage in tasks of significant relevance to the broad scope of the computer science community. Recent advances indicate that hybrid classical-quantum models can readily attain competitive performances at low architecture complexities. Such investigations are often carried out for image-processing tasks, and are notably constrained to modelling \textit{raster images}, represented as a grid of two-dimensional pixels. Here, we introduce vector-based representation of sketch drawings as a test-bed for QML models. Such a lower-dimensional data structure results handful to benchmark model's performance, particularly in current transition times, where classical simulations of quantum circuits are naturally limited in the number of qubits, and quantum hardware is not readily available to perform large-scale experiments. We report some encouraging results for primitive hybrid classical-quantum architectures, in a canonical sketch recognition problem.
\end{abstract}

\section{Introduction}~\label{sec:introduction}

Quantum information processing~\cite{nielsen2000book, Wilde2013book} holds the promise to be a game-changer in a wide variety of scenarios. Examples are ---but not restricted to--- providing energy-savings in large-scale computations~\cite{Fellous2023optimizing,Stevens2022energetics}, speeding-up linear-algebra subroutines~\cite{Harrow2009HHL}, boosting data-base search-routines~\cite{Grover1996fast,Giri2017reviewsearch} with applications in optimization problems~\cite{abbas2023optimization}, factorizing prime-numbers in polynomial time~\cite{shor1997polynomial}, breaking commonly-used cryptography protocols~\cite{Pirandola2020cripto}, and enhancing precision for sensing applications~\cite{Giovanetti2006Qmetro} among a constantly growing number of quantum algorithm applications~\cite{Dalzell2023survey}. In turn, experimental quantum-advantage has already been attained in specific scenarios related to learning quantum data~\cite{Huang2022Experimental}, sampling from very complex distributions~\cite{Zhou2022Experimental,Arute2019supremacy,Madsen2022computational}, or proof-of-concepts quantum-speeding-ups in reinforcement-learning~\cite{Saggio2021Experimental}, and this collection of results constitute a milestone in quantum hardware development. Nonetheless, there is still a long way to go until the aforementioned game-changer quantum algorithms can be implemented in real quantum hardware, since this requires a large amount of logical (\textit{e.g.} error-corrected) qubits~\footnote{We note, nevertheless, that the field is moving fast, and early fault-tolerant quantum computing is actively gaining protagonism~\cite{Bluvstein2024, liang2024modeling,kiss2024early}.}.

Driven by the advent of Artificial Intelligence, Quantum Machine Learning (QML)~\cite{djunko2016qml,Biamonte2017QML,Dunjko2018QAIrev,  Amin2018Qboltzmann, Khoshaman2019QVAE, krenn2023artificial, Cerezo2021, NISQreviewAlba, Dawid2022modern, Romero2017qautoencoder} has emerged as an appealing bridge between two rapidly-growing and highly impactful fields, holding a considerable attention from both scientific and industry communities. In this context, one of the most prominent fields that can potentially benefit from QML is Computer Vision. As a matter of fact, the massive computing resources required to train deep architectures, along with the huge amount of data needed to reach state-of-the-art accuracies, poses the question of whether alternative computing ways might provide a certain saving. Indeed, the possibility of analyzing visual data by a quantum model is increasingly gaining attention. Here, image-data is encoded into a quantum system, \textit{e.g.} quantum circuits or channels, thus allowing for quantum information processing in an exponentially large space. In turn, the so-called quantum image processing field has recently witness a notorious progress both from a theoretical but also experimental side~\cite{Haque2023,yan2020quantum,Lisnichenko2022,Wei2017qip,Defienne2022} (see also Refs.~\cite{tsang2016theory,tsang2020semi,tsang2019resolving,fiderer2021general,karsa2023illumination, pereira2023cluster, banchi2020barcode,paniate2023light,llorens2024qedg} for a more quantum-information oriented approach).

Despite lacking large-scale error-corrected quantum computers, a legitimate research question is whether currently-available technology, coined Noisy Intermediate-Scale Quantum (NISQ) devices~\cite{Preskill2018quantumcomputingin}, can be exploited for machine-learning and computer-vision applications. Here, not only the possibility of processing a larger amount of data is to be studied~\cite{Harrow2020small}, but also potential reductions in learning-models sizes~\cite{Jerbi2023beyond,schuld2020cc, Jerbi2021enhancements}, enhanced generalization capacities~\cite{Gyurik2023structuralrisk,Caro2022generalization, Caro2023generalization_bounds}, improved memory and adaptive capabilities~\cite{Elliot2022Qadaptive}, or its usage as natural platforms for generative-learning~\cite{Rudolph2023trainability,Innocenti2023quantum,HibatAllah2024Framework,Gili_2023doqbmg} are among the many NISQ applications recently proposed by the community.

Herein, Parameterized Quantum Circuits (PQCs) emerged as route to make use of NISQ devices. Variational Quantum Algorithms (VQAs)~\cite{peruzzo2014variational,Cerezo2021, NISQreviewAlba} train such circuits to minimize a cost function and consequently accomplish a task of interest; exampĺes are solving linear systems of equations~\cite{bravo2020variational,huang2019near,xu2019variational}, factoring~\cite{anschuetz2019variational}, analyzing principle components~\cite{larose2019variational} or devising parameterized quantum policies for reinforcement-learning applications~\cite{Skolik2022quantumagentsingym, meyer2022surveyQRL}. While relatively large-scale implementations of VQAs have been experimentally carried out~\cite{harrigan2021quantum,arute2020hartree,ollitrault2020quantum}, fundamental aspects related to the so-called barren-plateaus (BPs) currently hold back VQAs applications. Such  phenomenon is related to trainability~\cite{mcclean2018barren,Larocca2024reviewbp}, expressibility~\cite{holmes2020barren}, and overall classical-simulability~\cite{cerezo2024does} of variational quantum circuits. Essentially, many different scenarios (such as random parameter initialization, presence of hardware noise, choice of quantum circuit layout or cost-function non-locality) showcase BPs, implying exponentially vanishing cost-function gradients as the quantum circuit dimension (number of qubits) increases~\footnote{
While some analogies can be drawn between BPs and the vanishing-gradient problem in classical neural networks, there are non-trivial differences. While the former one is a genuinly quantum effect, the classical vanishing gradient problem depends on network's depth and not in the number of \textit{input networks}; we refer the interested reader to Sec.X of Ref.~\cite{Larocca2024reviewbp}
}. While the overall BP theory constitute a severe warning to the scalability of standard VQA approaches, hope in exploiting NISQ hardware is certainly not lost. In turn, a tremendous effort is being put forward to analyze alternative and non-standard methods for VQA applications, such as warm parameter initializations~\cite{puig2024warm,grant2019initialization}, variable-structure architectures~\cite{Bilkis2023vans,fosel2021quantum,economu2024testadaptvqe, patel2024curriculum, verdon2019learning}, data re-uploading techniques~\cite{PerezSalinas2020datareuploading}, and alternative optimization methods that hinge on quantum correlations~\cite{Burak2022Training,Liao2021quantum} or randomized measurements (\textit{e.g.} classical shadows~\cite{Huang2020predicting, huang2024certifying}).

In this context, the potential of hybrid classical-quantum architectures is increasingly gaining attention~\cite{domingo2023hybrid,sedykh2024hybrid,Kordzanganeh2023hybrid,Perelshtein2022PracticalAA, Chao2022bert, qi2021transfer,cantori2024synergy}. Such approaches relief NISQ computers from the task of fully processing data, but instead exploit deep-learning models capabilities to map the data into a convenient lower-dimensional manifold. Relevant features are subsequently post-processed by a NISQ device, that generally provides the output of the hybrid classical-quantum model. We remark that, as recently pointed out in Ref.~\cite{bowles2024better}, little is known on whether genuine quantum effects are actually employed in such hybrid approaches. Thus, we shall understand the field of hybrid classical-quantum machine learning as a highly exploratory one, that could potentially benefit from the synergies between the two fields in the mid-term, rather than providing a framework for practical quantum advantage in the short-term.

In this work, we analyze the potential of QML for sketch processing tasks. Most notably, previous approaches to quantum image processing focus in \textit{raster} images, \textit{e.g.} traditional pixel-based data. While such is a natural way to encode visual information, there are notwithstanding alternative ones. In turn, raster images encode data in a low-dimensional grid of pixels, which carries an overhead for machine-learning algorithms when performing feature-decoding, since the information is spread all over the grid in an unstructured way. From an human-level learning perspective, visual data is rarely conceptualized as a raster image. Rather, we represent our environment in terms of abstract concepts, \textit{e.g.} symbols or sketches~\cite{lake2015human,Yaroslav2018synthetizing, kulkarnani2015advances,das2020beziersketch}.

Here, we aim at providing a two-folded contribution. On the one hand, we introduce state-of-the-art hybrid classical-quantum QML methods to the broad computer science audience, encouraging this community to explore such an alternative way to do computations. On the other hand, our aim is also that of introducing sketch processing tasks to the QML community, stressing the opportunities and challenges that this computer vision problem has to offer.

\section{Framework}\label{sec:framework}
In the following we provide a short introduction to quantum machine learning (QML), and consequently to the field of sketch processing. Readers unfamiliar with quantum physics are encouraged to first go through Appendix~\ref{sec:qphys}, where basic notions such as quantum states, observables and measurements are discussed.

\subsection{Quantum Machine Learning}\label{ssec:qml}
Quantum models (QMs) exploit effects such as quantum superposition or entanglement in order to make computations in a probabilistic way. While universal and provably more powerful than classical computing~\cite{nielsen2000book,aaransoin2022howmuch}, we will here conceive them as special-purpose processing units, and restrict the following discussion to noiseless variational quantum computing; more interested readers are encouraged to follow Refs.~\cite{Cerezo2021, NISQreviewAlba} for further details.

\paragraph{Variational Quantum Algorithms.}VQAs make use of Parametrized Quantum Circuits (PQCs) to prepare target quantum states. Here, a fiducial $n$-qubit quantum state, often $\ket{0}^{\otimes n}$, gets transformed by a PQC, consisting on a series of quantum gates. Such quantum gates are chosen from a pre-defined dictionary (\textit{e.g.} single-qubit rotations and CNOTs), and placed at different positions in the circuit. We remark that devising mechanisms to build quantum-circuit layouts that can readily be adapted to current NISQ hardware and tackle trainability challenges is a topic of current research~\cite{Bilkis2023vans,fosel2021quantum,economu2024testadaptvqe, patel2024curriculum, verdon2019learning,diaz2023showcasing}.

In analogy to classical neural networks scenarios, the goal of VQAs is to minimize a given cost function $C(\vec{\theta})$, depending on \textit{trainable} parameter values appearing in the quantum circuit, which we denote as $\vec{\theta}$, and are usually associated to the single-qubit rotations. Somewhat misleading, this setting is often known as a \textit{quantum neural network}.

Here, a quantum device is used to estimate cost-function and cost-gradient values, whereas a classical optimization algorithm, \textit{e.g.} Adam optimizer, is in charge of navigating the parameter landscape. In its most basic format, the VQA training mechanism implies solving an optimization problem encoded into the cost-function structure as per the expectation value of a quantum observable $\mathcal{O}$:
\begin{equation}\label{eq:cost}
    C(\vec{x};\vec{\theta})= \langle \psi_{\vec{\theta}}(\vec{x})| \mathcal{O} |\psi_{\vec{\theta}}(\vec{x})\rangle,
\end{equation}
where $|\psi_{\vec{\theta}}(\vec{x})\rangle$ is the quantum state prepared by the PQC. Here, we remark the potential dependence of the quantum state in classical data $\vec{x}$, which can be injected by an appropriate encoding strategy. Note that such data $\vec{x}$ can in turn be the output of a classical machine-learning algorithm, as discussed in Sec.~\ref{sec:method}.

The estimation of the cost via the PQC, out of running it many times (so to accurately estimate its expected value) and its modification by its gradient, defines a VQA-cycle, which is iterated until meeting a stopping criteria~\footnote{Usually, the available measurements consist on local projections over the eigenbasis of $\sigma^{i}_z$ for each qubit $i=1,...,n$. In order to estimate the expected value of an arbitrary observable $\mathcal{O}$, we decompose such operator in terms of Pauli operators (which form a basis in the space of hermitian operators) and estimate each expected value by applying a proper change-of-basis transformation before measuring over the eigenbasis of $\sigma^{i}_z$ for each qubit. Thus, the cost-function can be estimated by computing averages of some pre-determined observables, which are in turn a weighted sum of expected values of Pauli-strings.}. We remark that cost-function gradients can be estimated via the so-called parameter-shift rule~\cite{Mitarai2018circuit,schuld2019evaluating,Wierichs2022generalparameter}. In short, this implies that for a broad class of cost-functions (as those considered in Eq.~\ref{eq:cost}), cost-function gradients can be estimated using the same circuit, but shifting the parameters according to a pre-defined \textit{recipe}. This is utmost importance for VQAs implementation on real quantum hardware, since it provides a route to compute gradients in an \textit{experimental} way. Alternatively, for classical simulation of quantum circuits (as done in this manuscript), quantum automatic-differentiation libraries have been developed (such as PennyLane~\cite{Bergholm2018pennylane}, TensorFlowQuantum~\cite{broughton2019tfq}, Qiskit~\cite{Qiskit} among others). As such, they are restricted in the number of qubits that can classically simulate, though provide a convenient framework to explore the potential interplay between state-of-the-art machine-learning algorithms and NISQ hardware.

The success of the VQA scheme hinges on several factors, and as discussed in Sec.~\ref{sec:introduction} serious trainability issues (as well as hardware noise) forbid the success of VQAs, constituting the main barrier to make use of current NISQ devices in the VQA framework. 

Recently, hybrid classical-quantum models have gained considerable attention. In this framework, the quantum circuit becomes an appendix of a (classical) deep-learning architecture. Such a setting has been studied for image classification~\cite{Senokosov_2024, Sagingalieva2023}, filtering~\cite{Shukla2022} and transfer-learning~\cite{Mari2020transferlearningin}, showing promising results in terms of architecture sizes. However, much work is still due in order to inject an inductive bias to the QML models, particularly when training on classical data~\cite{bowles2024better,Larocca2022group}. In this direction, we will now discuss sketch processing, a field that has not been tackled from a quantum computing perspective before.

\subsection{Sketch Processing}\label{ssec:sketch_processing}

Unlike pre-trained language models like GPT-k~\cite{brown2020language} or BERT~\cite{devlin2018bert}, humans can effortlessly interpret diverse and abstract conceptual sketches by leveraging our robust mental models, which integrate perceptual processing and high-level cognitive functions~\cite{tversky2002sketches}. This ability allows us to understand even the most ambiguous sketches, a skill that current AI systems struggle to match. Existing works~\cite{lake2015human, feinman2020learning} have tried to bridge the gap between human-like reasoning and AI models (which directly learn from raw data) by using neuro-symbolic models that capture compositionality and causal knowledge for flexible generalization. Other works~\cite{Ha2018,graves2013generating} have developed autoregressive approaches, utilizing similar stroke primitives to model the causal representation of sketches. By combining a classical autoregressive model of sketches with an integrated quantum processing module, we seek to explore a new spectrum of vector sketch modelling.

\textbf{Learning Sketch Representations.} Just as language comprises a set of syntactic rules and semantic structures, sketches can be interpreted as visual sentences~\cite{ganin2021computer,mas2010syntactic}, constructed through a series of pen strokes or \textit{words} that follow a certain grammar. In this proposed approach, an image is treated as a sequence of lines and represented as a \textit{vector} image. This analogy drives us towards modelling sketches in a sequential nature, where the drawings parallel the linear construction of sentences in natural language~\cite{tiwari2024sketchgpt}. The most commonly adopted format for a vector image in the classical sketch literature~\cite{Ha2018} is a sequence $\mathcal{D}$, each containing a 2D canvas coordinate $X_i$ sampled from a continuous drawing flow and a pen-state flag bit $f_i$ denoting whether the pen touches the canvas or not.
The vectorized sketch drawing $\mathcal{D}$ can be represented as:
\begin{equation}\label{eq:vector}
\mathcal{D} = \{(X_i, f_i)\}_{i=1}^L,
\end{equation}
where $X_i = (x_i, y_i )^T \in \mathbb{R}^2$, \( f_i \in \{\text{0}, \text{1}\} \) depending on whether the pen is up or down, and \( L \) denotes the length of the whole sketch.

In particular, SketchRNN~\cite{Ha2018} learns a parametric Recurrent Neural Network (RNN) to model the joint probability distribution $\mathcal{Q}$ of 2D coordinates and pen state as a product of conditionals as
\begin{equation}\label{eq:cond}
\mathcal{Q}(\mathcal{D}; \theta) = \prod_{i=1}^L p(X_i, f_i \mid X_{<i}, f_{<i}; \theta),
\end{equation}
where \( \theta \) is the set of learnt parameters of the model, and \( X_{<i} \) and \( f_{<i} \) denote the list of canvas locations and pen-state bits respectively before \( X_i \) and \( q_i \).

\textbf{Vectorized Sketch Data Processor:} SketchRNN treats
sketches as a digitized sequence of 2D points on a drawing canvas sampled along the ink-flow trajectory, \textit{e.g.} a sequence of dense line segments. This representation strategy has several drawbacks. In this sense, SketchRNN provides limited scalability, since it still models sketches as a sequence of (line-connected) pixels. As such, it is limited in processing long stroke sequences before the memory of the underlying RNN collapses.

\begin{figure}[t!]
    \includegraphics[width=\linewidth]{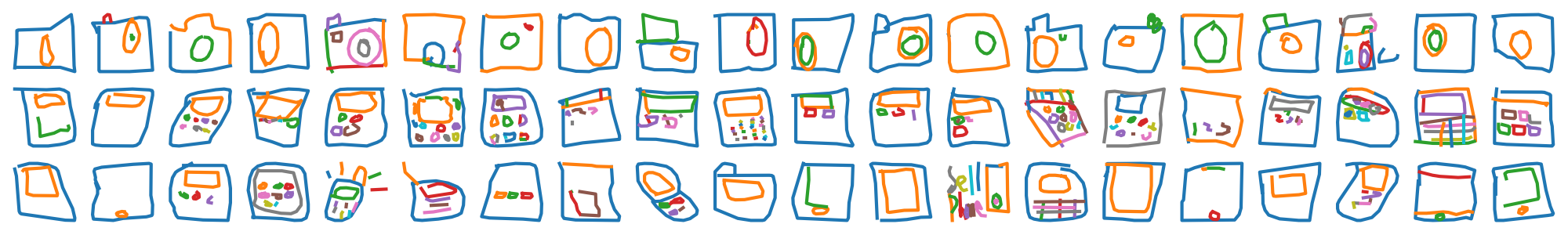}
    \caption{\textbf{Sketch Samples from QuickDraw Dataset}: This figure displays a curated selection of sketch drawings from the QuickDraw dataset. Each row represents a distinct category: cameras in the top row, calculators in the middle row, and cellphones in the bottom row. These examples illustrate the variety of drawing styles and levels of detail present in the dataset.}
    \label{fig:sketchsamples}
\end{figure}

Fig.~\ref{fig:sketchsamples} shows the complexity of the different sketch samples having variable curved lines and complex shapes representing a distinct class category from the QuickDraw dataset~\cite{jongejan2016quick}. These samples often carry significant fine-grained detail that straight lines cannot replicate. In a two-dimensional plane, drawing a straight line is straightforward: we simply connect two points with the shortest path. However, constructing a learnable sketch representation able to overcome the issues raised above requires simplifying stroke primitives in an abstract way, considering angles and twists times among other quantities~\cite{alaniz2022abstracting}.

In this work, we utilize parameterized \textit{Bezier curves} to encode human sketches~\cite{das2020beziersketch}. Bezier curves are parametrized polynomials that provide a smooth and scalable representation of a finite length curve, using a few control points. Here, \textit{anchor points} are fixed points between which the curve is drawn, and \textit{curvature points} are parameters that regulate the curvature intensity. The number of these points determines the \textit{order of the curve} (\textit{e.g.} a linear curve has two anchor points, a quadratic curve has one control point and two anchor points, while a cubic curve has two control points and two anchor points curves).

Hence, we here devise a stroke-to-Bezier encoding approach inspired from BezierSketch~\cite{das2020beziersketch} as

\begin{equation}
\left[\overrightarrow{\mathbf{s}_i}, \overleftarrow{\mathbf{s}_i}\right]=\mathcal{\phi}\left(\mathbf{X}_{i-1}, \mathbf{f}_{i-1} ; \theta\right)
\label{eq:encoding}
\end{equation}

\begin{equation}
\mathcal{B}=\mathbf{W}_{\mathcal{B}}\left[\overrightarrow{\mathbf{s}}_{e n d} ; \overleftarrow{\mathbf{s}}_{e n d}\right],
\end{equation}

where $\mathcal{B}$ denotes the Bezier-fit representation extracted from the last hidden state extracted from forward and backward states $\overrightarrow{\mathbf{s}}_{e n d}$ and $\overleftarrow{\mathbf{s}}_{e n d}$ respectively; the interested reader is referred to Ref.~\cite{das2020beziersketch} for further details on this embedding.

\paragraph{Other Sketch Applications.} The mechanism described above allows for a plethora of different applications, and are  due to the introduction of the TU-Berlin~\cite{eitz2010sketch}, Sketchy~\cite{sangkloy2016sketchy} and QuickDraw datasets. Among the applications, we aim to highlight sketch recognition and generation tasks. Here, the introduction of large-scale sketch datasets~\cite{jongejan2016quick,sangkloy2016sketchy} paved the way for deep learning models~\cite{song2017deep,yu2016sketch}, which has achieved performance surpassing that of humans. Other application domains of interest include sketch-based image retrieval~\cite{dey2019doodle}, creative sketch generation~\cite{bhunia2022doodleformer}, sketch-based object localization~\cite{tripathi2020sketch} where abstract sketches serve as conceptual representations.

\section{Methodology}\label{sec:method}
In this work we approach the task of understanding abstract sketches from a QML angle. The search of quantum advantage, one of the most visited topics for quantum computing, is beyond the scope of this manuscript; our intention is to help elucidating novel research directions, related to learning how to conceptualize out of quantum machines by analyzing symbolic data~\cite{lake2015human}.

Here, we take first steps towards that direction by tackling the sketch recognition problem. This relies more on causal and compositional conceptual reasoning rather than classification tasks where models directly learn from pixel-based raw image data. In the latter case, the MNIST data-set~\cite{lecun-mnisthandwrittendigit-2010} is used in a strikingly frequent way. This standard dataset is composed of handwritten digits and currently used a standard benchmark for QML models. However, it consists on $60K$ pixel-based images ($28$ x $28$), which exceeds the dimension of data that can be handled by either currently available quantum hardware (or classical simulators), if we do not rely on dimensionality reductions and constrain the number of labels to be classified, posing serious limitations to benchmark QML models. This issue points out the necessity for the QML community to explore different tasks and associated datasets whose samples are  diverse in the way in which information is coded. In the following sections we will use of the QuickDraw dataset as an example of this approach within the task of sketch recognition.

\textbf{Choice of Dataset.}
We curated a subset of the QuickDraw dataset that comprises three classes. We specifically selected images of cellphones, cameras, and calculators from the QuickDraw. These classes are chosen due to their visual similarities, which present a greater challenge for classification tasks. By focusing on these closely related categories, we aim to push the boundaries of our model's performance and gain deeper insights into its ability to distinguish between similar objects.The rationale behind this decision can be based on the practical need to optimise computing resources at the time of simulating large data-sets on quantum circuits.  This approach not only tests the robustness of our model but also mimics real-world scenarios where such distinctions are crucial. Note that working with a subset of the QuickDraw dataset implies the necessity of defining a new baseline, a matter discussed in Sec.~\ref{ssec:baseline}.

\subsection{QuantumDraw Framework}\label{ssec:qdraw}
\begin{figure}[t!]
    \includegraphics[width=\linewidth]{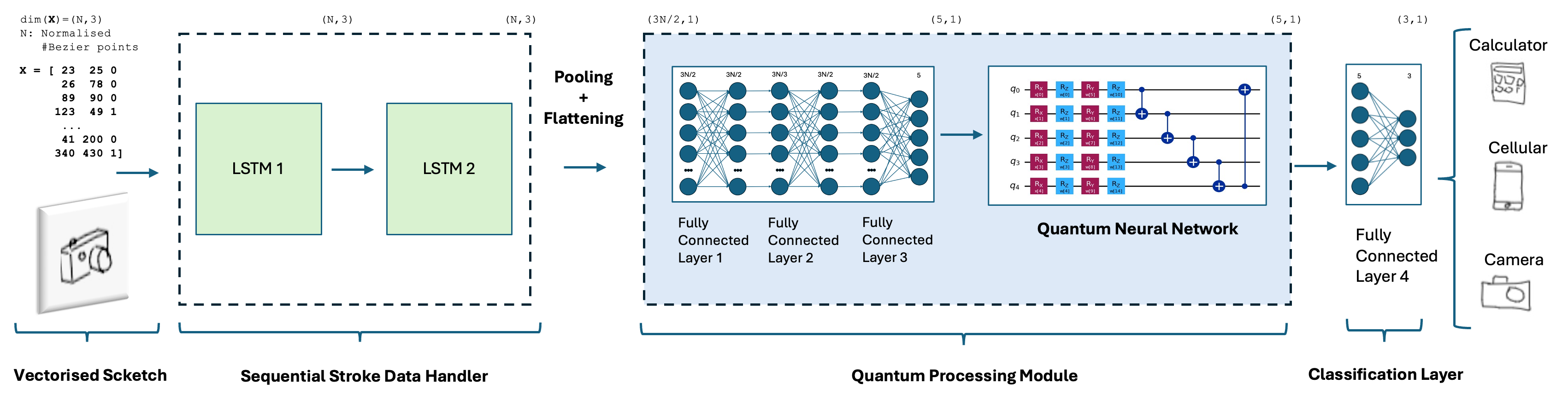}
    \caption{\textbf{Overview of QuantumDraw Framework}: The vectorised drawing include the Bezier point coordinates and a flag for end-of-stroke. Sequential Stroke Data Handler is constituted by two LSTM units. Data is half-sized through max-pooling, and flattened to a vector feeding the Quantum Processing Module, with three fully connected layers, ending in five cells to feed the quantum circuit. A final fully connected layer provides the classification.}
    \label{fig:system}
\end{figure}

The overview of our hybrid classical-quantum model is illustrated in Fig.~\ref{fig:system}, and consists on the following three building blocks: the Sequential Stroke Data Handler, the Quantum Processing Module and the Final Classification Layer. Sketches are vectorised in a set containing the Bezier point coordinates for every stroke, together with a flag for end-of-stroke. In order to keep the length of every sample uniform, samples (sketches) are normalised to the maximum number of points in the database, which we denote by $N$, padding with zeros to keep constant $N$-length. No time-stamps are used for this work.

\noindent
\textbf{Sequential-Stroke Data Handler:}
The Sequential-Stroke Data Handler consists of two Long Short-Term Memory (LSTM) units~\cite{hochreiter1997long}, chosen to build a sequential-encoder block that mimics Sketch-RNN~\cite{Ha2018}. Such recurrent layers encode the input vector image-sequence $\mathcal{D}$, and lead to a sequence of hidden states $h_i$. The final hidden state is further post-processed by means of a max-pool operation (that halves the size of the data).

\noindent
\textbf{Quantum Processing Module:}
In order to make feasible the hybrid classical-quantum approach, some sort of dimensionality reduction is required~\cite{domingo2023hybrid,sedykh2024hybrid,Kordzanganeh2023hybrid,Perelshtein2022PracticalAA, Chao2022bert, qi2021transfer,cantori2024synergy}. Here, we consider three fully-connected neural networks whose output is fed to a quantum circuit. The fully connected networks are two-layer networks, the initial pair keeping the original dimensionality of the hidden state output by the Sequential Stroke Data Handler (the recurrent module), and the final one ending in five output networks, which are fed to a five-qubit quantum circuit as rotation angles of $R_x$ gates acting on each qubit.

We chose a standard quantum circuit layout, known as Hardware Efficient Anstaz (HEA)~\cite{peruzzo2014variational}, and use a single layer (classical network outputs are fed by an angle-embedding in the first single-qubit rotation acting on each qubit). We remark that there is room to improve on such circuit layout (\textit{e.g.} by a variable-ansatz approach~\cite{Bilkis2023vans,fosel2021quantum,economu2024testadaptvqe, patel2024curriculum, verdon2019learning}), and our choice is aligned with keeping the architectures as simple as possible. The HEA consists of three consecutive single-qubit rotations $(R_y, R_z, R_y)$ followed by a CNOT gate; rotations are parametrized by real numbers defining rotation angles, which here serve as the trainable parameters of the parametrized quantum circuit. CNOTs, on the other hand, act on two qubits, and generate quantum correlations between them; this differentiates classical to quantum computing paradigm, being thus of utmost relevance. We note that a dictionary comprised of single-qubit rotations and CNOTs connecting all qubits in the circuit is approximately universal~\cite{nielsen2000book}: any target $n$-qubit quantum transformation can be implemented provided enough gates of the dictionary are used. However, issues arise since when stacking many HEA-layers together, the circuit becomes too expressible and BPs rise~\cite{holmes2020barren}.

\noindent
\textbf{Final classifier:} A final fully-connected neural network layer provides the classification. This is done by feeding the five expected-values that the quantum circuit computes ($\langle \sigma^{(i)}_z \rangle$) for $i=1,...,5$ to the classical network, resulting in three numbers describing the probability that the original sketch belong to each class-type (in our case: \textit{Calculator}, \textit{Cellular} and \textit{Camera}).

\subsection{Classical Baseline}\label{ssec:baseline}
As a baseline, we adopt a simple RNN encoder consisting of a couple of LSTM layer, followed by a feed-forward layer that outputs the probability for each category. The primary aim of using a simple RNN model is not to push the boundaries of state-of-the-art performance but rather to set a foundational benchmark for analysis. This allows us to focus our discussion on the conceptual insights of our findings rather than developing another highly optimized model.

\subsection{Implementation Details}
The QuickDraw dataset is initially pre-processed to normalize the images, ensuring a consistent input shape suitable for our model. The data is divided into training and validation sets to facilitate the evaluation of the model's performance during training. We use the Adam method as our optimizer, which adapts the learning rate for each parameter, and cross entropy as our objective function, which is well-suited for recognition tasks. The model is trained over a series of $100$ epochs, during which the optimizer updates the model weights to minimize the loss function. Throughout the training, both training and validation losses are monitored. Moreover, we perform experiments on multiple seeds to assess model's learning progress and generalization ability.

\section{Results}\label{sec:results}
We present results for the sketch-processing problem by means of hybrid classical-quantum models. All the experiments and results can be found in open access through to the code provided as supporting material to this piece of research \footnote{Y.Cordero, “Qdraw”: https://github.com/yeray142/QDraw. Open source code and data.}.

We train our classical baseline on the three classes considered of the vectorized-image QuickDraw dataset. Next, we train our QuantumDraw model. As detailed in Sec.~\ref{ssec:qdraw}, our model is an hybrid architecture that first process the full sketch by two consecutive LSTM layers. The final hidden state is reduced in dimensionality and provides the input to a feed-forward network, whose output is in turn processed by a five-qubit quantum circuit. Finally, the output of the five expected values provided by the quantum circuit are post-processed by a single-layered feed-forward network that retrieves the probability for the sketch belonging to each category.

Additionally, and for comparative purposes, we analyze the same dataset, but now by keeping each sample encoded as a raster-image (instead of a vectorised one). To this end, we implemented the Hybrid Quantum Neural Network-Parallel (HQNN-Parallel) circuit, proposed in Ref.~\cite{Senokosov_2024}. Such a model consists on a series of (classical) convolutional layers that extract the relevant features of the (pixel-based) image to reduce the original dimensionality. These features are then processed by parallel parametrized quantum circuits, whose outputs (expected values) are further post-processed by a feed-forward layer retrieving the probability of the sketch belonging to a given category.

\begin{figure}[t!]
  \centering
  \includegraphics[width=0.32\textwidth]{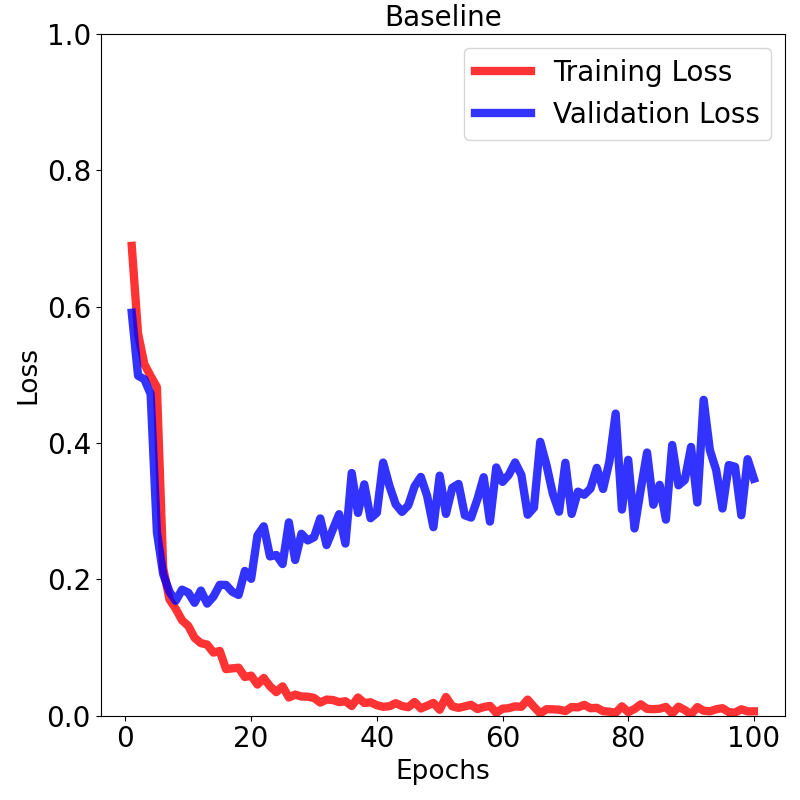}
  \includegraphics[width=0.32\linewidth]{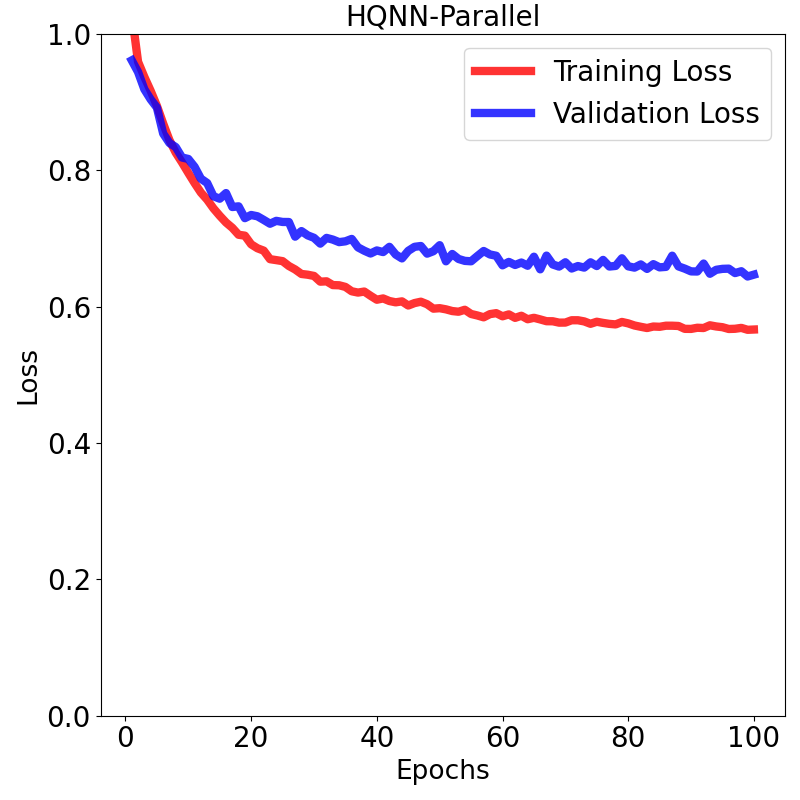}
    \includegraphics[width=0.32\linewidth]{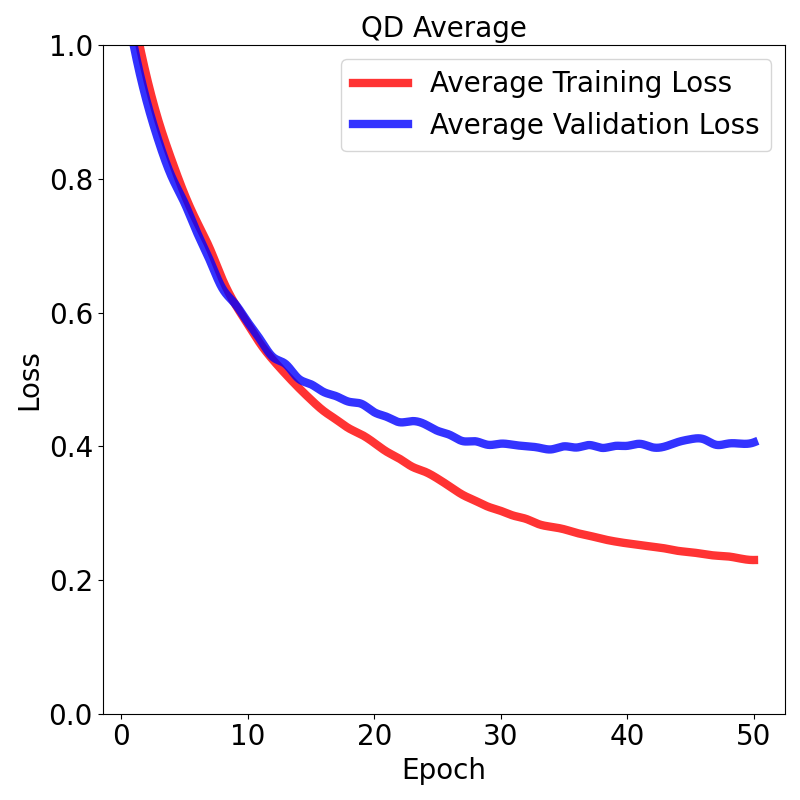}
  \caption{\textbf{Comparative Analysis of the QuantumDraw Model}: We show training and validation performance on \textit{(left)} baseline model consisting on two LSTM layers plus feed-forward post-processing, \textit{(center)} raster-image classification carried out by the HQNN-Parallel circuit (see main body), and \textit{(right)} vectorized-image classificaton carried out by our QuantumDraw model, explained in Sec.~\ref{ssec:qdraw}.}
  \label{fig:benchmark}
\end{figure}

 The learning curves for each model are depicted in Fig.~\ref{fig:benchmark}, while Table \ref{tab:comparison} shows the accuracy results.

\subsection{Discussion} Our results indicate that our QuantumDraw model is able to capture complex patterns and generalize effectively to unseen data, with a performance comparable to that of its classical baseline. However, care must be taken when claiming such a competitive results, and in the following we further analyze the \textit{quantumness} associated to the proposed model and its resulted trained version.

To shed light into this result, we studied two different versions of QuantumDraw, whose learning curves are reported in Appendix~\ref{sec:apb}. First, we considered QD-Frozen, a version of QuantumDraw in which the trainable (quantum) parameters are frozen to their (initially randomly chosen) value; here the classical weights of the LSTM need to adapt in order to counteract the quantum layer, which is not being trained. The aim of such a scheme is to test whether the classical part of the model can compensate the presence of the quantum circuit. Our results point in the opposite direction, indicating that training both the classical \textit{and} the quantum networks leads to an improved performance, as compared to trianing only the classical part.

As pointed out in the introduction, little is known on whether genuine quantum effects are actually employed in hybrid approaches as the one proposed in the this piece of research. For instance, even though there is strong theoretical basis to include quantum correlations as core ingredient on the design of hybrid classical-quantum systems, and a growing number of bibliographic contributions are supporting its implementation in a diverse number of applications \cite{bowles2024better}, this is still a highly exploratory field, and it is worth to analyse to what extent the presence or absence of the entanglement are actually impacting the final result. For this reason, we consider a second (separable) version of the QuantumDraw, deemed QD-Sep, in which the CNOTs present in the quantum circuit layout are removed. By doing this, we remove quantum correlations present in the circuit, and thus the model can be simulated classically in an efficient way. We trained many several instances of QD-Sep and obtained competitive performances than its original version with quantum correlations. This result, which is in strong alignment with current warnings of the overall trainability and classically simulatibility of quantum models~\cite{cerezo2024does, bowles2024better} is to be taken with care.

Our hybrid model is to be understood as a platform to allow further explorations, rather than providing a quantum advantage. In this sense, we did not tackle in this paper the optimization of the quantum circuit layout to take into account potential symmetries and advantages of the sketch structure and recognition problem, nor to fully benefit from the intrinsic temporality of the data. Rather than ruling out any room for quantum advantage in frameworks similar to our QuantumDraw, the QD-Sep results should be understood --- we believe --- as a message to both QML community and computer science community. When searching for concrete applications of quantum technology, room for basic science and negative results should be allowed. The path that scientific knowledge has historically taken is erratic, and we shall not expect that straightfowardly implementing ``quantum" analogs of neural networks can lead to an improvement over state-of-the art mechanisms.

\begin{table}[t!]
  \centering
  \begin{tabular}{lcccc}
    \hline
    Model & Training Accuracy & Validation Accuracy & Training Loss & Validation Loss \\
    \hline
    QD & 0.89 (0.67, 0.99) & 0.85 (0.61, 0.96) & 0.31 (0.07, 0.46) & 0.44 (0.20, 0.58) \\
    QD-Frozen & 0.66 & 0.65 & 0.40 & 0.56 \\
    QD-Sep & 0.93 (0.66, 0.99) & 0.87 (0.59, 0.93) & 0.22 (0.03, 0.41) & \textbf{0.39} (0.26, 0.65) \\
    HQNN-Parallel & \textbf{0.99} & \textbf{0.91} & 0.57 & 0.65 \\
    Classical baseline & 0.94 & 0.76 & \textbf{0.18} & 0.60 \\
    \hline
  \end{tabular}
  \caption{We compare training and validation metrics for QuantumDraw (QD), QD-sep, QD-Frozen, HQNN-Parallel and classical baseline. Accuracy refers to the percentage of correctly recognized sketches, indicating the model's classification performance. Loss, measured using cross-entropy loss, indicates the error in sketch classification. The Adam optimizer was employed during the training process. The classic model shows the highest accuracy and lowest loss values. When evaluating the performance over several instances, we report average, minimum, and maximum values.}
  \label{tab:comparison}
\end{table}

Table~\ref{tab:comparison} presents the comparison of training and validation accuracies and losses for the various models studied in our experiments. It also includes the results for the classical model, as well as several configurations of the quantum models, including average, minimum, maximum, frozen, and separable parameter settings.

\section{Conclusion and Future Scopes}\label{sec:future_work}

In this paper, we have shown how to tackle the sketch-processing problem from the alternative paradigm of quantum computing, initiating this field of research.

Rather than inspecting quantum advantages, our goal is to bring two distant communities closer, since we believe that interesting synergies can emerge from such approach. In particular, we analyzed the possibility of processing vector images by means of a hybrid classical-quantum model for a three-class sketch-recognition problem. While further investigations are required, and in particular increasing the number of classes (at a considerably high computational cost), our results are encouraging, and further investigations need to be carried out in order to reduce the variance of our results and analyze whether the baseline considered can be simplified further.

From a research direction path, this work opens up the possibility to \textit{import} many canonical sketch-processing tasks to the QML realm. This is not constrained to recognition, but also generation, completion and abstraction among others~\cite{lake2015human}. However, we remark that to do this, a different QML architecture is to be considered, since partial sketch information needs to be accessed in order to predict the next stroke. This is an ongoing investigation, and it is an open question whether quantum memory effects can help in such auto-regression or not.

While the quantum processing of classical information is still at early stages, and little is known on how to cleverly approach it, we believe that computer vision problems such as sketch classification can serve as good test-beds, precisely for the reasons mentioned above. In contrast, most of the QML literature tests their models on raster images, more often the MNIST dataset~\cite{lecun-mnisthandwrittendigit-2010}, a dataset made requiring relevant computational power to train even on hybrid classical-quantum models~\cite{Senokosov_2024}. Moreover, images are represented as a two-dimensional pixel grid, which needs to be down-scaled in order to fit into either simulable quantum devices or near-term quantum hardware.  On the contrary, vector-based images are not only of lower dimension, but also present an interesting and largely unexplored temporal nature. In turn, quantum-enhanced memory agents can potentially be beneficial at the time of processing sketches.

As we mentioned along the paper, the database used for the sketches include the timestamp for every stroke. Time stamps were not strictly included in the analysis carried out on this paper, and only the end-of-stroke flag was integrated into the image representation. However, this can be considered as a promising future line of work, by conditioning the quantum model with temporal dependencies provided by the available time stamp information.

Finally, we believe that much of the discussions hold by the computer vision community can be beneficial in order to re-consider different aspects of quantum machine learning, not necessarily constrained to classical-data processing. In this sense, it would be interesting to study how conceptualizations of (quantum) data can be done by a quantum computer, a matter that will ultimately lead us to better understand human reasoning.

\section*{Acknowledgments}
We specially thanks Lic. Tomás Crosta for helpful discussions regarding this work. M.B. acknowledges support from AGAUR Grant no. 2023 INV-2 00034 funded by the European Union, Next Generation EU and Grant PID2021-126808OB-I00 funded by MCIN/AEI/ 10.13039/501100011033 and by ERDF A way of making Europe. F.V and M.B acknowledge the support from the Spanish Ministry of Science and Innovation through the project GRAIL, grant no. PID291-1268080B-100 and the Cátedra UAB-Cruïlla, TSI-100929-2023-2.
\medskip

{
\small
\bibliographystyle{ieeetr}

}

\newpage
\appendix
\section{Selected basic concepts of quantum physics and quantum information}\label{sec:qphys}

The smallest quantum system is deemed a \textit{qubit}, and for composite $n$-qubit systems we have $d=2^n$~\footnote{Our discussion is here restricted to finite-dimensional systems, but we remark that infinite-dimensional, known as continuous variable systems, also play a fundamental role in quantum machine learning~\cite{killoran2019cv,weedbrook2012gaussian}}. This poses a serious limitation to classical simulations of quantum systems, due to exponentially-growing memory requirements, being \textit{brute-force} approaches constrained to $\leq 32$ qubits~\cite{xu2023heculean}.

A quantum system is fully characterized by its quantum state, represented by a $d$-dimensional complex vector
\begin{equation}\label{eq:psi}
\ket{\psi} = \sum_{k=0}^{d-1}c_k \ket{k},
\end{equation}
whose squared-absolute entry values sum up to one, \textit{i.e.} $\sum_k |c_k|^2=1$. Here, $\llaves{\ket{k}}$ denotes an orthononormal basis in a $d$-dimensional Hilbert space, and $c_k$ are complex coefficients denoting amplitudes of quantum state $\ket{\psi}$. Note we are here using the so-called \textit{Dirac} notation, denoting vector $\vec{k}\equiv\ket{k}$.

The way to extract information out of a quantum system is by \textit{measuring} it; here we are interested in \textit{projective measurements} (which are only a sub-class of the whole set of quantum measurements), and are obtained by projecting the quantum state into proper sub-spaces of an \textit{observable} $\mathcal{K}$ of interest. The latter denotes a physical quantity (\textit{e.g.} position, angular-momentum, hamiltonian) and is mathematically represented by an hermitian operator, implying its equipped with an associated eigenbasis $\llaves{\ket{k}}$. In particular, a projective measurement on $\ket{\psi}$ over $\llaves{\ket{k}}$ retrieves a \textit{measurement outcome} $k$ with probability $p(k|\ket{\psi}) = |c_k|^2$. This provides an operational interpretation of the coefficients $c_k$ in Eq.~\eqref{eq:psi}. Importantly, when the quantum system is (projectively) measured, it is also \textit{destroyed}, and the state $\ket{\psi}$ needs to be prepared again; for a detailed discussion about quantum measurements we refer the reader to Refs.~\cite{nielsen2000book, Wiseman_Milburn_2009}.

As an example, a (pure) state of a qubit can be parametrized as per $\ket{\psi} = \cos(\frac{\theta}{2}) \ket{0} + e^{-i \phi}\sin(\frac{\theta}{2})  \ket{1}$, with $\llaves{\ket{0}, \ket{1}}$ being the eigenstates of $\sigma_z$, \textit{i.e.} the Pauli matrix representing the observable \textit{intrinsic angular momentum} of a spin-$1/2$ particle in the $z$ direction. Quantum states are connected by unitary transformations, which in the single-qubit case are given by rotations; $(\theta,\phi)$ are parameters associated to rotation-angles. By appropriately choosing the rotation-angles, we can reach any target state. While this idea generalizes to bigger systems, the exponential growth of Hilbert space make finding the adequate transformation a highly non-trivial task.

In turn, the Hilbert space associated to composite $n$-qubit systems is given by the direct product of individual single-qubit Hilbert spaces. As a consequence, the direct product of local eigenbasis spans a global eigenbasis as $\llaves{k} = \llaves{\ket{0},\ket{1}}^{\otimes n}$. The latter is in correspondence to the observable $\bigoplus^n\sigma_z^{(i)}$, where $\sigma_z^{(i)}$ is the spin-$z$ operator of the $i^{\text{th}}$ qubit (up to a constant factor). This example is relevant in gate-based quantum computing, since such is the quantum measurement usually performed after preparing the quantum state~\footnote{Note that a different observables can be measured by applying a suitable change-of-basis transformation.}

\section{Additional experiment details}\label{sec:apb}
\begin{figure}[t!]
  \centering
        \includegraphics[width=1.\linewidth]{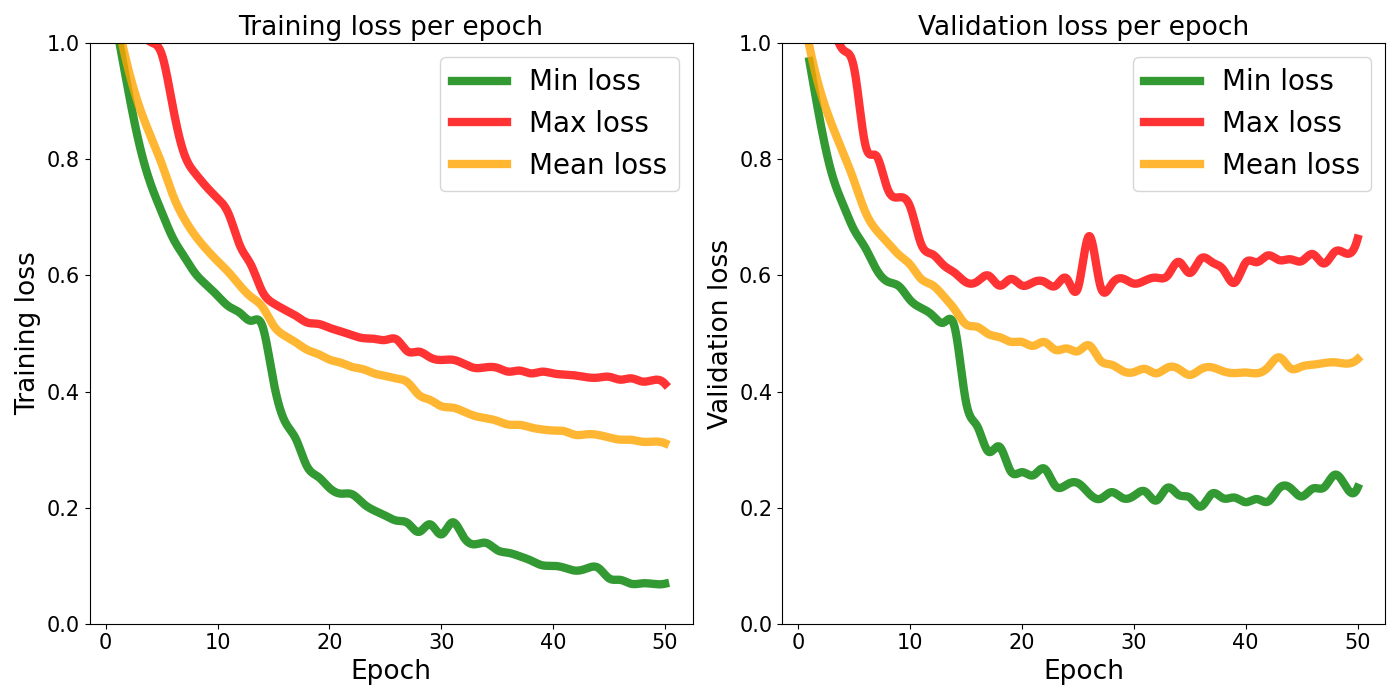}
  \caption{\textit{Average QuantumDraw performance}. We show learning curves of our hybrid model, initialized over 50 different seeds. We show training/validation (left/right) loss per epoch, along with minimum, maximum, and mean training losses. The validation dataset consists of a set of data points not seen by the model during training, used to evaluate the model's ability to generalize to new, unseen data. }
  \label{fig:qraw_avg}
\end{figure}

We report further numerical analysis the QuantumDraw model behaviour. To begin with, we report its average learning curve, when (trainable) quantum paramters are randomly initialized over different seeds (see Fig.~\ref{fig:qraw_avg}).

\begin{figure}[t!]
  \centering
        \includegraphics[width=.45\linewidth]{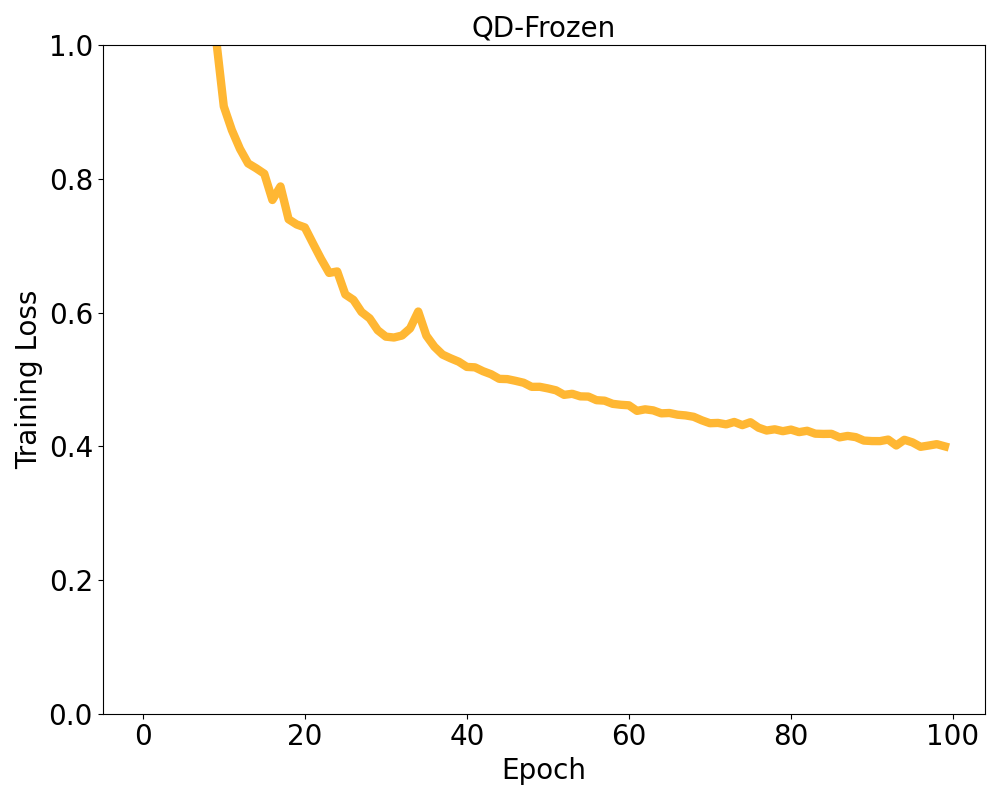}
        \includegraphics[width=.45\linewidth]{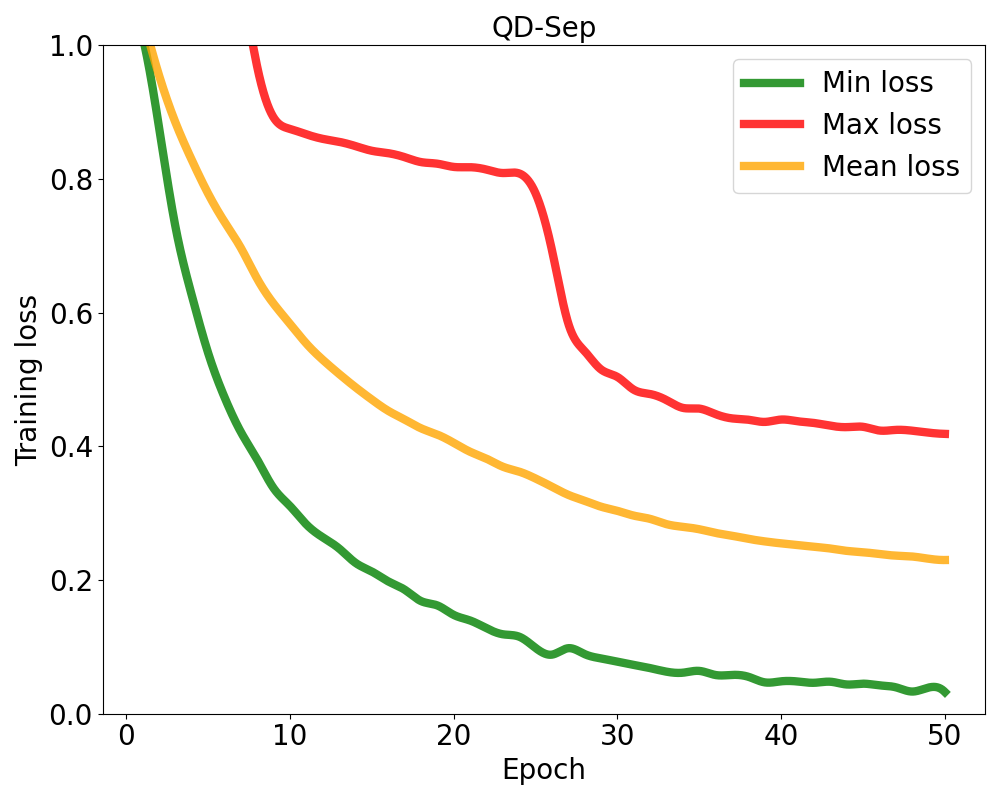}
  \caption{\textit{Quantumness testing}. We show learning curves for two models that test the quantumness/classicality of the QuantumDraw model. \textit{(Left):} QD-frozen quantum parameters associated to rotation-angles are randomly initialized and not modified during training, while the classical part of the model needs to adapt its weights to optimize the cost value. \textit{(Right:)} we show the QD-Sep model, in which CNOTs are removed from the quantum circuit layout.}
  \label{fig:separable_frozen}
\end{figure}

Next, we study the case in which the quantum trainable parameters are \textit{freezed}, while the classical weights of the network are allowed to be modified; learning curves are shown in Fig.~\ref{fig:separable_frozen} (left). Here, we want to understand whether the quantum circuit, plays a relevant role or not, and if its presence can be \textit{overcomed} by adapting only modifying classical weights. Our results indicate that such is not the case (see Table~\ref{tab:comparison}), since the performance of the freezed quantum model (QM-Frozen) is worst.

Finally, we consider the case in which the full hybrid QuantumDraw model is trained, but the entangling gates are removed, \textit{e.g.} making it \textit{separable}; learning curves are shown in Fig.~\ref{fig:separable_frozen} (right). This makes the quantum model classically simulable, and thus serves as a test of model's \textit{quantumness}~\cite{bowles2024better}. In turn, and in strong alignement with Ref.~\cite{bowles2024better} and current results from the QML community, the hybrid classical-quantum models do not actually seem to require quantum entanglement to achieve their best performances, and thus they can actually be classically simulable in an efficient way.

\end{document}